\documentclass[reprint, superscriptaddress, nofootinbib, prd]{revtex4-1}

\usepackage{amsmath}
\usepackage{amssymb}
\usepackage{graphicx}
\usepackage[utf8]{inputenc}
\usepackage{verbatim}
\usepackage[T1]{fontenc}
\usepackage[english]{babel}
\usepackage{dcolumn}
\usepackage{bm}
\usepackage{mathrsfs}  
\usepackage{mathtools}
\usepackage{cancel}
\usepackage{esvect}
\usepackage{subfigure}
\usepackage{booktabs}
\usepackage{soul}
\usepackage{bbold}
\usepackage[dvipsnames]{xcolor}
\usepackage{braket}

\definecolor{custompurple}{HTML}{9300d3}
\definecolor{customgreen}{HTML}{019d73}
\definecolor{customblue}{HTML}{57b5e8}

\usepackage[colorlinks=true,linkcolor=customblue, citecolor=blue,urlcolor=blue]{hyperref}
\newcommand{\tr}[1]{\mathrm{Tr}\left[#1\right]}

\newcommand{\dif}{\mathrm{d}}
\newcommand{\PB}[1]{\left\{#1\right\}_{\text{PB}}}
\newcommand{\DB}[1]{\left\{#1\right\}_{\text{DB}}}

\begin{document}

\preprint{}
\title{Covariant equations of motion of massive spinning particles in a background Yang-Mills field}
\author{Jie Zhou}
\affiliation{State Key Laboratory of Dark Matter Physics, Shanghai Key Laboratory for Particle Physics and Cosmology,
Key Laboratory for Particle Astrophysics and Cosmology (MOE),
School of Physics and Astronomy, Shanghai Jiao Tong University, Shanghai 200240, China}
\author{Ying Shan Zhao}
\affiliation{State Key Laboratory of Dark Matter Physics, Shanghai Key Laboratory for Particle Physics and Cosmology,
Key Laboratory for Particle Astrophysics and Cosmology (MOE),
School of Physics and Astronomy, Shanghai Jiao Tong University, Shanghai 200240, China}
\author{Yifeng Sun}
\email{Contact author: sunyfphy@sjtu.edu.cn}
\affiliation{State Key Laboratory of Dark Matter Physics, Shanghai Key Laboratory for Particle Physics and Cosmology,
Key Laboratory for Particle Astrophysics and Cosmology (MOE),
School of Physics and Astronomy, Shanghai Jiao Tong University, Shanghai 200240, China}

\date{\today}

\begin{abstract}
The dynamics of a spinning colored particle in a background non-Abelian Yang-Mills field is of broad interest in many areas of physics. A physically important application arises in relativistic heavy-ion collisions, where hard probes such as heavy quarks and jets propagate through the strong early-time classical color fields collectively referred to as the glasma. The standard framework for describing the classical dynamics of colored particles in a background Yang-Mills field is provided by the Wong equations, but it does not incorporate spin degrees of freedom. Although several extensions of the Wong equations have been proposed to include spin, they generally fail to satisfy all the necessary requirements simultaneously, such as Lorentz covariance, allowance for an arbitrary chromomagnetic moment, and preservation of the required physical constraints. In this work, we extend the framework of a relativistic classical spinning particle in an electromagnetic field to describe spin-1/2 quarks propagating in a generic background non-Abelian Yang-Mills field. By systematically applying the Dirac-Bergmann algorithm, we derive a self-consistent set of equations of motion for the particle’s coordinates, momenta, spin, and color charge that satisfies all these requirements. This formalism provides a more complete and physically consistent description of spinning colored particles in background Yang-Mills fields, and offers a suitable framework for studying momentum diffusion and spin polarization phenomena of hard probes in heavy-ion collisions, particularly in the glasma.
\end{abstract}

\maketitle


\section{Introduction}

Relativistic heavy-ion collision experiments conducted at the Large Hadron Collider (LHC)~\cite{Aamodt:2008zz} and the Relativistic Heavy Ion Collider (RHIC)~\cite{Adams:2005dq,Adcox:2004mh,Arsene:2004fa,Back:2004je} provide a unique opportunity to create and study a new state of matter composed of deconfined quarks and gluons, known as the quark-gluon plasma (QGP). According to the color glass condensate (CGC) effective theory~\cite{Iancu:2003xm,Gelis:2010nm,Gelis:2012ri}, immediately after the collision and prior to the formation of the QGP, the medium exhibits large gluon occupation numbers and can be described by classical color fields, collectively referred to as the glasma~\cite{Lappi:2006fp,Lappi:2008eq,Fujii:2008km,Fukushima:2011nq}. Hard probes, such as heavy quarks (charm and bottom) and energetic jets, are produced with a characteristic formation time of $\tau_0\le0.1\,\mathrm{fm}/c$ in heavy-ion collisions. These probes interact strongly with the glasma fields and thus provide valuable information about the properties and structure of the initially produced matter. Numerous studies have demonstrated that the glasma can significantly influence the momentum broadening of heavy quarks and jets, as well as the angular correlations between heavy-flavor particles~\cite{Das:2015aga,Das:2017dsh,Ruggieri:2018rzi,Sun:2019fud,Liu:2019lac,Liu:2020cpj,Khowal:2021zoo,Ruggieri:2022kxv,Ipp:2020mjc,Ipp:2020nfu,Avramescu:2023qvv,Avramescu:2024xts}.

In noncentral relativistic heavy-ion collisions, a strong vorticity field is generated in the QGP as a result of the large orbital angular momentum deposited in the overlap region. This vorticity field can induce global polarization phenomena of spinning hadrons, such as $\Lambda$ hyperons, and spin alignment of spin-one mesons through spin-orbital coupling~\cite{Liang_2005,Gao:2007bc}. Both global and local spin polarization, as well as spin alignment, have been observed by the STAR Collaboration~\cite{STAR:2017ckg,Adam_2019,STAR:2022fan} and the ALICE Collaboration~\cite{ALICE:2019aid,Acharya_2020,ALICE:2022dyy}. Although the underlying mechanisms responsible for the local spin polarization and spin alignment are not yet fully understood~\cite{Sun:2018bjl,Liu:2019krs,Fu:2021pok,Becattini:2021iol,Sheng:2022wsy,Li:2022vmb}, studies of these spin phenomena open new avenues for exploring the dynamics and properties of the QGP created in relativistic heavy-ion collisions. Since hard probes carry both spin and color and interact with the glasma field in the early stage of relativistic heavy-ion collisions, it has been proposed in Refs.~\cite{Muller:2021hpe,Kumar:2023ghs,Yang:2024qpy} that strong glasma fields may also induce spin polarization and spin alignment phenomena. Therefore, heavy quarks and jets serve as sensitive probes of the glasma, not only through their momentum distributions but also via their spin degrees of freedom. 

Most existing studies of glasma effects on heavy quark and jet transport rely on Wong equations~\cite{Wong:1970fu}, which describe the evolution of coordinate, kinetic momentum, and color charge in a classical non-Abelian field. Since the glasma field is highly non-uniform, such approaches may overlook important effects, for example, the Stern-Gerlach force acting on spinning particles in inhomogeneous fields. This motivates the derivation of equations of motion for spinning colored particles in a generic background non-Abelian Yang-Mills field, a topic of broad interest in several areas of physics.
Several attempts have been made to extend Wong equations by incorporating spin degrees of freedom; however, these formulations exhibit various shortcomings. In Refs.~\cite{Balachandran:1976ya,Mueller:2019gjj}, the equations of motion for spin and color were constructed using Grassmann variables, which makes it difficult to describe spin at the semiclassical level~\cite{Deriglazov_2014}. The approaches in Refs.~\cite{Heinz:1984my,Yang:2021fea}, derived respectively from the quadratic form of the QCD Dirac equation and from the Wigner-function formalism with an $\hbar$ expansion, cannot simultaneously accommodate an arbitrary chromomagnetic moment while preserving all the necessary physical constraints, such as the mass-shell condition and the Tulczyjew-Dixon spin supplementary condition~\cite{Tulczyjew1959,Dixon:1970zza}. 
We note that Ref.~\cite{Linden_1996} provides a nonrelativistic formulation of spinning particles in background Yang-Mills fields, and Refs.~\cite{Ghosh:1994hn,Ghosh:1994jq} develop consistent Lagrangian and Hamiltonian descriptions of spinning anyons in 2+1 dimensions; however, these frameworks cannot be applied to the highly relativistic particles propagating in the Yang-Mills fields.
\par In order to consistently describe spin and color dynamics simultaneously in the case of an arbitrary chromomagnetic moment $g_s\neq2$ and in a generic background non-Abelian Yang-Mills field, we build on the constrained Hamiltonian framework developed in Refs.~\cite{DERIGLAZOV_2013,Deriglazov:2013zaa,Deriglazov_2014} for spin-1/2 particles in electromagnetic backgrounds, which consistently incorporates the necessary physical constraints. A key feature of the present work is that it extends the constraint-consistent Hamiltonian description of relativistic spinning particles to the non-Abelian case, so that the non-Abelian color degrees of freedom, their Lie-Poisson structure, and the associated physical constraints are incorporated consistently within the same dynamical framework. Within this new constrained Hamiltonian framework, we derive a self-consistent set of equations of motion that are manifestly Lorentz covariant and compatible with the physical constraints associated with momentum, spin and color charge. We also present explicitly the Dirac brackets among the dynamical variables in the presence of a background non-Abelian Yang-Mills field, which define the constrained phase-space structure of the system, underlie the derivation of the equations of motion, and naturally determine the phase-space measure relevant for kinetic theory. In addition, we explicitly establish two structural properties that are essential in the non-Abelian case, namely, the conservation of the classical Casimir invariants of the color charge and the gauge covariance of the equations of motion. These equations provide a unified basis for studying transport coefficients and spin-related phenomena of heavy quarks in background non-Abelian fields. 

\par This paper is organized as follows. In Sec.~\ref{se:cd}, we introduce the dynamical variables, their constraints, and the fundamental Poisson relations. Section~\ref{se:cHM} provides an overview of the Hamiltonian formalism for constrained systems. In Sec.~\ref{se:eom}, we employ this formalism to derive the equations of motion for a quark propagating in a background Yang-Mills field. Finally, Sec.~\ref{se:dis} presents a summary and discussion of our results.

\par Throughout this work, we use natural units with $\hbar = c = 1$. The metric tensor of Minkowski space is defined as $g^{\mu\nu} = \mathrm{diag}(+1, -1, -1, -1)$, and the Levi-Civita tensor conventions are $\epsilon^{0123} = +1$ and $\epsilon_{0123} = -1$. The Einstein summation convention is implied for repeated indices, and repeated color indices within an expression are also understood to be summed.

\section{Dynamical variables, constraints, and basic Poisson relations}
\label{se:cd}

The classical equations of motion of a point like particle can be formulated either in the Lagrangian or Hamiltonian framework. In the Lagrangian approach, one specifies the explicit form of the Lagrangian as a function of the dynamical variables and their time derivatives. In the Hamiltonian approach, one instead defines the phase-space variables and the corresponding Poisson brackets among them. To ensure that the resulting equations of motion rigorously satisfy all physical constraints, we employ the Hamiltonian formalism in this study.

To describe the dynamics of a colored, spinning point like particle in a background Yang-Mills field while maintaining Lorentz covariance, we introduce the following dynamical variables: the four-vectors $x^\mu$ and $P^\mu$ representing the canonical coordinate and kinetic momentum, respectively; the color charge $q^a$; and the antisymmetric tensor $S^{\mu\nu}$ representing the spin. The total number of degrees of freedom is therefore $2 \times 4 = 8$ for the canonical coordinates and kinetic momenta, $N_c^2 - 1$ for the color charge (where $N_c$ denotes the number of colors in the $\mathrm{SU}(N_c)$ Yang-Mills gauge theory), and 6 for the spin tensor. However, these variables are not all independent. After imposing the physical constraints, the number of independent dynamical degrees of freedom is reduced to $6+N_c^2-N_c+2$. Specifically, the coordinate-momentum sector contributes 6 degrees of freedom, the color sector contributes $N^2_c-N_c$ because the $N_c^2-1$ color variables are subject to $N_c-1$ Casimir constraints, and the spin sector contributes 2 physical degrees of freedom corresponding to a spin-1/2 particle.

\subsection{Canonical coordinate and momentum}

For a nonrelativistic point like particle without spin or color charge, the standard Hamiltonian is expressed as a function of the canonical coordinates and momenta $(\boldsymbol{x}, \boldsymbol{p})$ in three-dimensional space. The only nonvanishing Poisson bracket between their components is
\begin{equation}
    \PB{x_i, p_j} = \delta_{ij}.
\end{equation}
To generalize this formulation to relativistic particles, one introduces the four-vector forms of the canonical coordinate $x^\mu$ and canonical momentum $p^\mu$. Accordingly, the Poisson bracket is extended, and the only nonvanishing relation becomes
\begin{equation}
    \PB{x^\mu, p^\nu} = g^{\mu\nu}.
    \label{eq_canon}
\end{equation}

The canonical momentum $p^\mu$ is related to the kinetic momentum $P^\mu$ through
\begin{equation}
    P^\mu = p^\mu - qe A^\mu,
    \label{eq:P}
\end{equation}
where $A^\mu$ denotes the background $\mathrm{U(1)}$ gauge potential, $e$ is the coupling constant, and $q$ is the charge of the particle. In the presence of a background non-Abelian Yang-Mills field, the kinetic momentum takes the form \begin{equation}
	P^\mu=p^\mu-gq^aA^{\mu,a},
\end{equation} 
where $A^{\mu,a}$ denotes the background $\mathrm{SU(N_c)}$ gauge potential, $g$ is the coupling constant, and $q^a$ is the  corresponding color charge of particle.

In the relativistic Hamiltonian framework, the on-shell condition for the kinetic momentum $P^\mu$ must be imposed as a constraint. For a spinless particle, the well-known mass-shell condition takes the form
\begin{equation}
    P^2 - m^2 = 0.
    \label{eq_Einstein}
\end{equation}
When spin degrees of freedom are included, the spin-orbit interaction modifies the on-shell condition. For a classical spin-$1/2$ particle in an external gauge field, the constraint differs from Eq.~(\ref{eq_Einstein}) and takes the widely adopted form~\cite{Heinz:1984my,Deriglazov_2014,Liu_2020}
\begin{equation}
    P^2 - \frac{\mu}{2} qe F^{\mu\nu} S_{\mu\nu} - m^2 = 0,
    \label{eq:onshell}
\end{equation}
where $F^{\mu\nu}$ is the field-strength tensor of the external gauge field, $m$ is the particle mass, and $\mu$ denotes the anomalous magnetic moment. The parameter $\mu$ is related to the Lande $g$ factor by $\mu = g_S / 2$, where $g_S$ need not be exactly 2 once quantum corrections are taken into account.

\subsection{Classical color charges of quarks} 

In quantum chromodynamics (QCD), the interaction between quarks and gluons is described by the Lagrangian term
\begin{equation}
    \mathcal{L}_{\text{int}} = g\, \bar{\psi} \gamma_{\mu} A^{\mu} \psi 
    = g\, \bar{\psi} \gamma_{\mu} T^{a} \psi\, A^{\mu,a},
    \label{eq_int}
\end{equation}
from which the color current of a quark can be identified as $j^{\mu,a} = \bar{\psi} \gamma^{\mu} T^{a} \psi$. 
Here, $T^a$ are the generators of the SU(3) color group in the fundamental representation, satisfying the commutation relation
\begin{equation}
    [T^a, T^b] = i f^{abc} T^c,
\end{equation}
where $f^{abc}$ are the structure constants of the SU(3) Lie algebra, and the color index $a$ runs from $1$ to $8$.

For a classical point like particle carrying color charge and propagating in a background SU($N_c$) Yang-Mills field, the corresponding color current can be expressed in terms of a real Lie algebra-valued variable $q^a$ as~\cite{Balasin_2014}
\begin{equation}
    j^{\mu,a}(y) = \int d\lambda\, q^a(\lambda)\, \dot{x}^\mu\, \delta^{(4)}(y - x(\lambda)),
\end{equation}
where $a = 1, \ldots, N_c^2 - 1$ and $\lambda$ is the worldline parameter. 
In this picture, $q^a$ serves as the classical color charge. 
From Eq.~(\ref{eq_int}), one may identify the quantum operator of color charge as $\hat{q}^a \equiv T^a$, i.e., the generators of SU($N_c$) acting on the quark color state.

The Hamiltonian formalism is formulated on phase space endowed with a Poisson-bracket structure. For the operator-valued color charges $\hat{q}^a$, the corresponding commutation relation reads
\begin{equation}
    [\hat{q}^a, \hat{q}^b] = i f^{abc} \hat{q}^c.
\end{equation}
Taking the classical limit, one replaces the commutator by the Poisson bracket, leading to~\cite{Johnson:1988qm,BULGAC1990187,Litim_2002}
\begin{equation}
    \label{eq:qP}
    \PB{q^a, q^b} = f^{abc} q^c.
\end{equation}

Equation~(\ref{eq:qP}) indicates that the classical color charges $q^a$ are not canonical variables. 
Since the conventional Hamiltonian approach relies on canonical coordinates, it is often convenient to introduce canonical (Darboux) variables that parametrize the color degrees of freedom. 
For SU($N_c$), one can define $N_c(N_c - 1)/2$ pairs of canonical variables $(\boldsymbol{\phi}, \boldsymbol{\xi})$ satisfying $\PB{\phi_i, \xi_j} = \delta_{ij}$. 
Explicit constructions for the SU(2) and SU(3) cases can be found in Refs.~\cite{Johnson:1988qm,BULGAC1990187,Litim_2002}.

However, in the present work we do not employ the Darboux parametrization of color charges. 
As pointed out in Ref.~\cite{BULGAC1990187}, such parametrizations are neither necessary nor globally valid, since any specific coordinate choice covers only a portion of the underlying group manifold, making it impossible to define canonical variables globally for SU($N_c$). 
Instead, as we demonstrate in Sec.~\ref{se:eom}, the color charge $q^a$ itself can be directly treated as a dynamical variable within the Hamiltonian framework. 
The Poisson bracket relation~(\ref{eq:qP}) is sufficient for deriving consistent equations of motion that satisfy the color charge constraints introduced below.

\par
For an SU($N_c$) gauge group, there exist $N_c - 1$ independent Casimir operators that commute with all generators $T^a$ and characterize the total color charge magnitude. 
In general, the Casimir operators of the SU($N_c$) group are defined as
\begin{equation}
    Q_n = d^{abc\ldots} T^a T^b T^c \ldots,
\end{equation}
where $d^{abc\ldots}$ is a totally symmetric tensor given by
\begin{equation}
    d^{abc\ldots} \propto \mathrm{Tr}\left( T^a T^b T^c \ldots \right)_{\text{sym}},
\end{equation}
and the subscript ``sym'' denotes complete symmetrization over all indices.

For SU(2), there is only one quadratic Casimir operator, while for SU(3) there are two independent Casimir operators: the quadratic and cubic Casimirs. 
They are defined respectively as
\begin{equation}
    \label{eq:c2}
    T^a T^a = I_{D_r} C_2,
\end{equation}
and
\begin{equation}
    \label{eq:c3}
    d^{abc} T^a T^b T^c = I_{D_r} C_3,
\end{equation}
where $d^{abc}$ denotes the symmetric structure constants of the SU($N_c$) group and $I_{D_r}$ is the identity operator in the representation space of dimension $D_r$. For the fundamental representation, the corresponding eigenvalues of the quadratic and cubic Casimir operators are
\begin{subequations}
    \begin{gather}
        C_2 = \dfrac{N_c^2 - 1}{2N_c}, \\
        C_3 = \dfrac{(N_c^2 - 4)(N_c^2 - 1)}{4N_c^2}.
    \end{gather}
\end{subequations}

In the classical limit, one can define the analogous Casimirs  in terms of the classical color charges $q^a$ as~\cite{Kelly:1994ig,Litim_2002}
\begin{equation}
    Q_n = d^{abc\ldots} q^a q^b q^c \ldots.
\end{equation}
In particular, the quadratic and cubic classical Casimirs are
\begin{subequations}
    \begin{gather}
        Q_2 = q^a q^a, \\
        Q_3 = d^{abc} q^a q^b q^c.
    \end{gather}
\end{subequations}
\par
At the purely classical level, these Casimirs are conserved quantities, but their values need not \textit{a priori} be fixed to those of any particular quantum representation. When the classical colored particle is interpreted as the semiclassical limit of a quark, it is natural to match them to those of the fundamental representation. Specifically, adopting the matching condition $Q_{2,3} = D_r C_{2,3}$ \cite{Avramescu:2023qvv}, the quadratic and cubic classical Casimirs take the explicit values
\begin{subequations}
    \begin{gather}
        Q_2 = \dfrac{N_c^2 - 1}{2}, \\
        Q_3 = \dfrac{(N_c^2 - 4)(N_c^2 - 1)}{4N_c}.
    \end{gather}
\end{subequations}

The Casimirs  $Q_n$ must remain constant during time evolution. 
Therefore, the classical color charges are required to satisfy the self-consistency condition
\begin{equation}
    \label{eq:Q23c}
    \dot{Q}_n = 0,
\end{equation}
for all Casimirs  $Q_n$.
\subsection{Classical spin}

The classical description of spin can be formulated in two equivalent ways: 
either as an antisymmetric rank-2 tensor $S^{\mu\nu}$, following Frenkel~\cite{Frenkel1926}, 
or as a four-vector $S^{\mu}$, following the work of Bargmann, Michel, and Telegdi (BMT)~\cite{PhysRevLett.2.435}. 
The spin tensor $S^{\mu\nu}$ has six components; however, only two of them correspond to independent physical degrees of freedom. 
The reduction is achieved by imposing the following constraints. 
First, the normalization condition
\begin{equation}
	S^{\mu\nu}S_{\mu\nu} = 8s^2,
	\label{eq:s2t}
\end{equation}
where $s=1/2$ for quarks. To remove the additional three redundant components of the antisymmetric spin tensor, one needs to impose a spin supplementary condition. Several such conditions have been discussed in the literature, such as the Tulczyjew-Dixon (TD) condition \cite{Tulczyjew1959, Dixon:1970zza}, the Frenkel-Mathisson-Pirani (FMP) condition \cite{Frenkel1926, Mathisson1937}, and the Corinaldesi-Papapetrou (CP) condition \cite{Corinaldesi1951}. In the present work, we adopt the Tulczyjew-Dixon condition because it is a manifestly covariant, momentum-based condition that is physically natural for relativistic spinning particles and is also the choice most consistent with the constrained Hamiltonian phase-space formulation used here:
\begin{equation}
	\label{eq:pst}
	P^\mu S_{\mu\nu} = 0.
\end{equation}

\par Since the spin space possesses two physical degrees of freedom, it is often convenient to introduce a BMT-type spin four-vector $S^{\mu}$, defined analogously to the Pauli-Lubanski vector but with a slightly modified normalization~\cite{Deriglazov_2014}:
\begin{equation}
	\label{eq:S}
	S^\mu = \dfrac{1}{2\sqrt{P^2}}\, \epsilon^{\mu\nu\alpha\beta} P_\nu S_{\alpha\beta}.
\end{equation}
This vector satisfies the constraints
\begin{subequations}
	\begin{gather}
		P^\mu S_\mu = 0,\\
		S^\mu S_\mu = -4s^2,
		\label{eq:spin}
	\end{gather}
\end{subequations}
which again ensure that only two independent spin degrees of freedom remain.
In the case of $P^2 \neq 0$ and $P^\mu S_\mu = 0$, one can invert Eq.~(\ref{eq:S}) to express the spin tensor in terms of the spin vector:
\begin{equation}
	S^{\mu\nu} = \dfrac{1}{\sqrt{P^2}}\, \epsilon^{\mu\nu\rho\sigma} P_\rho S_\sigma.
\end{equation}

The spin four-vector and spin tensor are not canonical variables in the spin phase space. 
To construct a canonical formulation, we follow Ref.~\cite{Deriglazov_2014} and introduce a pair of canonical variables $(\omega^\mu, \pi^\nu)$, where $\omega^\mu$ and $\pi^\nu$ denote the  spin canonical coordinate and conjugate momentum, respectively. 
They satisfy the canonical Poisson brackets
\begin{equation}
	\PB{\omega^\mu, \pi^\nu} = g^{\mu\nu}.
	\label{eq_spin}
\end{equation}
The spin tensor can then be expressed in terms of these variables as~\cite{Deriglazov_2014}
\begin{equation}
	\label{eq:St}
	S^{\mu\nu} = 2(\omega^\mu \pi^\nu - \omega^\nu \pi^\mu).
\end{equation}
The Poisson brackets for the classical spin tensor are then given by~\cite{Deriglazov:2011wx}
\begin{equation}
\label{PBspin}
\PB{S^{\mu\nu},\, S^{\rho\sigma}}
= 2(
g^{\nu\rho} S^{\mu\sigma}
- g^{\mu\rho} S^{\nu\sigma}
+ g^{\mu\sigma} S^{\nu\rho}
- g^{\nu\sigma} S^{\mu\rho}),
\end{equation}
which provides the classical analog of the quantum commutation relation for the generators of the SO(1,3) Lorentz group,
\begin{equation}
[J^{\mu\nu},\, J^{\rho\sigma}]
= i(
g^{\nu\rho} J^{\mu\sigma}
- g^{\mu\rho} J^{\nu\sigma}
+ g^{\mu\sigma} J^{\nu\rho}
- g^{\nu\sigma} J^{\mu\rho}).
\end{equation}
The factor of 2 in Eq.~(\ref{PBspin}) arises from the normalization convention chosen for the spin tensor, as indicated in Eq.~(\ref{eq:s2t}) or (\ref{eq:spin}).

When considering a classical color-charged particle interacting with a background non-Abelian Yang-Mills field, 
the on-shell constraint for the kinetic momentum must be modified:
\begin{equation}
	P^2 - 2\mu gq^a F^{a}_{\mu\nu}\, \omega^\mu \pi^\nu - m^2 = 0,
	\label{eq_shell}
\end{equation}
where $\mu$ denotes the chromomagnetic moment of the particle, and $P_{\mu} = p_{\mu} - g q^{a} A_{\mu}^{a}$.

Regarding the constraints, Eqs.~(\ref{eq:s2t}) and (\ref{eq:pst}) imply
\begin{subequations}
    \begin{gather}
        \label{eq:Sp}
        S^{\mu\nu}S_{\mu\nu} = 8\big[\omega^2 \pi^2 - (\omega^\mu \pi_\mu)^2\big] = 8s^2,\\
        \label{eq:PSp}
        P^\mu \omega_\mu = 0, \quad P^\mu \pi_\mu = 0.
    \end{gather}
\end{subequations}
To further restrict the dynamics of the spin variables on the SO(1,3)-invariant spin surface, we impose additional constraints on $(\omega^\mu, \pi^\nu)$:
\begin{equation}
    \omega^2 = a_3, \quad \pi^2 = a_2, \quad \omega^\mu \pi_\mu = 0,
    \label{eq_normal}
\end{equation}
where $a_2$ and $a_3$ are constants whose specific values are not essential, as they do not affect the physical dynamics.

With these elements established, we proceed to derive the equations of motion for a classical colored and spinning particle interacting with a background non-Abelian Yang-Mills field. 
The full set of dynamical variables is $\zeta = (x^\mu, P^\mu, \omega^\mu, \pi^\nu, q^a)$, 
the constraints are given by Eqs.~(\ref{eq:Q23c}), (\ref{eq_shell}), (\ref{eq:PSp}), and (\ref{eq_normal}), 
and the basic non-vanishing Poisson brackets are Eqs.~(\ref{eq_canon}), (\ref{eq:qP}), and (\ref{eq_spin}).  

Using these fundamental Poisson brackets, one can write a compact expression for the Poisson bracket between any two functions $F$ and $G$ of the phase space variables $\zeta$:
\begin{equation}
    \PB{F,G} = \frac{\partial F}{\partial \zeta_i} \frac{\partial G}{\partial \zeta_j} \PB{\zeta_i, \zeta_j},
    \label{eq:Poisson}
\end{equation}
which explicitly yields
\begin{equation}
    \label{eq:PBd}
    \begin{split}
        \PB{F,G} &= \frac{\partial F}{\partial x_\mu} \frac{\partial G}{\partial p^\mu} - \frac{\partial F}{\partial p_\mu} \frac{\partial G}{\partial x^\mu} 
        + \frac{\partial F}{\partial \omega_\mu} \frac{\partial G}{\partial \pi^\mu} - \frac{\partial F}{\partial \pi_\mu} \frac{\partial G}{\partial \omega^\mu} \\
        &\quad + f^{abc} q^c \frac{\partial F}{\partial q^a} \frac{\partial G}{\partial q^b}.
    \end{split}
\end{equation}

The remaining task is to construct a Hamiltonian that generates equations of motion which rigorously respect the constraints (\ref{eq:Q23c}), (\ref{eq_shell}), (\ref{eq:PSp}), and (\ref{eq_normal}). 
The standard approach for such constrained Hamiltonian systems was developed in the seminal works of Dirac~\cite{Dirac1950,Dirac1951,Dirac1958a} and Bergmann~\cite{Anderson:1951ta,Bergmann:1954tc}, who established a consistent formalism, now known as the Dirac-Bergmann algorithm. 
This algorithm provides the framework used in the present study.

\section{Hamiltonian formalism for constrained systems}
\label{se:cHM}

A dynamical system can be described either in the Lagrangian or the Hamiltonian formalism, which are related by a Legendre transformation. 
Given a Lagrangian $L(q,\dot{q})$ expressed in terms of generalized coordinates $q_i$ and velocities $\dot{q}_i$, the equations of motion can be obtained from the Euler-Lagrange equations. 
In the Hamiltonian formalism, one defines the canonical momenta 
\begin{equation}
    p_i = \frac{\partial L}{\partial \dot{q}_i},
\end{equation}
then rewrites the velocities in terms of $(q,p)$ when possible, and defines the canonical Hamiltonian as
\begin{equation}
    H_C = \sum_i p_i \dot{q}_i - L.
\end{equation}
The equations of motion for canonical coordinates and momenta are then given by
\begin{equation}
    \dot{q}_i = \PB{q_i, H_C}, \quad \dot{p}_i = \PB{p_i, H_C}.
\end{equation}

However, in some cases the Hessian matrix of the Lagrangian
\begin{equation}
    W_{ij} = \frac{\partial^2 L}{\partial \dot{q}_i \partial \dot{q}_j}
\end{equation}
fails to be invertible, i.e., $\det(W) = 0$. Such a Lagrangian is called \emph{singular}, which implies that not all accelerations can be solved uniquely in terms of coordinates and velocities. Equivalently, one cannot express all $\dot{q}_i$ in terms of $(q,p)$, and this imposes constraints among the canonical variables,
\begin{equation}
    \phi_a(p,q) = 0,
\end{equation}
called \emph{primary constraints}. The Dirac-Bergmann algorithm provides a systematic procedure to handle such singular Lagrangian systems and convert them into a consistent constrained Hamiltonian formalism.

For a constrained system, the canonical Hamiltonian $H_C$ can always be expressed as a function of $(q,p)$, even if the velocities cannot be inverted. However, $H_C$ alone does not generate the correct dynamics corresponding to the original Lagrangian. Instead, the \emph{primary Hamiltonian}, which correctly reproduces the Lagrangian equations of motion, is defined as
\begin{equation}
    H_P \equiv H_C + \lambda_a \phi_a,
\end{equation}
where $\phi_a$ denotes the $a$-th primary constraint and $\lambda_a$ are undetermined Lagrange multipliers enforcing these constraints. The time evolution of any function $F$ on the phase space, including canonical variables and constraints, is then governed by the Poisson bracket with $H_P$:
\begin{align}
    \dot{F} &= \PB{F, H_P} = \PB{F, H_C} + \lambda_a \PB{F, \phi_a} \nonumber \\
    & + \phi_a \PB{F, \lambda_a}\approx \PB{F, H_C} + \lambda_a \PB{F, \phi_a},
\end{align}
where the symbol $\approx$ denotes \emph{weak equality}, i.e., equality on the constrained phase space.

Since primary constraints generally depend on the canonical variables, their time evolution must preserve the constraints, leading to the consistency condition
\begin{equation}
    \dot{\phi}_a = \PB{\phi_a, H_P} \approx 0.
\end{equation}
This condition may generate new constraints or impose restrictions on the Lagrange multipliers $\lambda_a$. 
The consistency procedure is iterated until no further constraints arise. 
All resulting constraints, including the primary as well as any secondary, tertiary, or higher-order constraints, define the allowed subspace of the phase space. 
For simplicity, we denote all constraints collectively as
\begin{equation}
    \phi_a = 0.
\end{equation}

The general solution for the Lagrange multipliers $\lambda_a$ can be inserted into the primary Hamiltonian $H_P$, yielding the \emph{total Hamiltonian} $H_T$:
\begin{equation}
    H_T = H_P \big|_{\text{inserting } \lambda_a}.
\end{equation}
In some cases, all $\lambda_a$ are determined, while in others, some Lagrange multipliers remain undetermined. The existence of under-determined multipliers is associated with the presence of \emph{first-class constraints}. A first-class constraint, denoted as $\chi_i$, has a vanishing Poisson bracket with all constraints:
\begin{equation}
    \PB{\chi_i, \phi_b} \approx 0,
\end{equation}
where $b$ runs over all constraints obtained from the consistency procedure.  

A practical method to identify first-class constraints is as follows. Construct the matrix
\begin{equation}
    \mathcal{C}_{ab} = \PB{\phi_a, \phi_b}.
\end{equation}
The null vectors $v_{i,a}$ of this matrix correspond to first-class constraints:
\begin{equation}
    \chi_i = v_{i,a} \phi_a.
\end{equation}
First-class constraints indicate the presence of gauge freedom. The associated under-determined Lagrange multipliers can either be left in $H_T$ or fixed by imposing gauge conditions, which convert these constraints into \emph{second-class constraints}, denoted as $\xi_\alpha$. Second-class constraints are those that have non-vanishing Poisson brackets with at least one other constraint.

The choice of $\xi_\alpha$ is not unique. In practice, one can select $N-R$ constraints from arbitrary combinations of the $\phi_a$, where $N$ is the total number of constraints and $R$ is the number of first-class constraints (equivalently, the rank of $\mathcal{C}$). As long as the matrix
\begin{equation}
    \Delta_{\alpha\beta} = \PB{\xi_\alpha, \xi_\beta}
\end{equation}
is invertible, i.e., $\det(\Delta) \neq 0$, the chosen $\xi_\alpha$ constitute a valid set of second-class constraints.

Since $H_T$ contains only the primary first-class constraints, while the Dirac-Bergmann algorithm may generate additional secondary (or higher-stage) first-class constraints, Dirac proposed enlarging the Hamiltonian to include all of them. This leads to the \emph{extended Hamiltonian},
\begin{equation}
    H_E = H_T + \lambda_i \chi_i,
    \label{eq_HE}
\end{equation}
where the $\chi_i$ denote all first-class constraints and $\lambda_i$ are arbitrary Lagrange multipliers.
With this extension, the time evolution of any phase-space function $F$ is governed by the Poisson bracket with $H_E$:
\begin{equation}
\label{eom_HE}
    \dot{F} = \PB{F,H_E}.
\end{equation}

To simplify the derivation of equations of motion, one introduces the \emph{Dirac bracket}:
\begin{equation}
    \label{eq:DBd}
    \DB{F, G} \equiv \PB{F, G} - \PB{F, \xi_\alpha} \Delta^{-1}_{\alpha\beta} \PB{\xi_\beta, G}.
\end{equation}
Since $H_E$ has vanishing Poisson brackets with all constraints on the constrained phase space,
\begin{equation}
    \PB{F, \xi_\alpha} \Delta^{-1}_{\alpha\beta} \PB{\xi_\beta, H_E} \approx 0,
\end{equation}
and thus on the constrained subspace
\begin{equation}
    \dot{F} = \PB{F, H_E} \approx \DB{F, H_E}.
    \label{eq_DB}
\end{equation}
All second-class constraints in $H_E$ can be eliminated via the Dirac bracket. Consequently, the equations of motion can equivalently be written in terms of the \emph{partially reduced Hamiltonian} $H_{PR}$:
\begin{equation}
    \dot{F} \approx \DB{F, H_{PR}},
    \label{eq_fr}
\end{equation}
where
\begin{equation}
    H_{PR} = H_E|_{\xi_\alpha = 0}.
\end{equation}
It is important to note that imposing second-class constraints in the first-class constraints and applying the Dirac bracket reproduces exactly the same first-class constraints. For further details on the Dirac-Bergmann algorithm, see Refs.~\cite{DiracBook,Wipf,Brown:2022gha}.

\section{Covariant equations of motion for spin-$1/2$ colored particles in a background non-Abelian Yang-Mills field}
\label{se:eom}

Applying the equation of motion~(\ref{eom_HE}) together with the Poisson-bracket relation~(\ref{eq:PBd}), 
the evolution equation for the color charge \(q = q^a T^a\) can be written as
\begin{equation}
    \dot{q} = [F, q],
\end{equation}
where \(F = -i\,(\partial H_E / \partial q^a)\, T^a\).
The time evolution of the classical Casimir invariants then follows:
\begin{align}
    \dot{Q}_n 
    &= n\, d^{abc\cdots}\, \dot{q}^a q^b q^c \cdots
       \propto n\, \mathrm{Tr}[\dot{q}\, q^{n-1}] 
       \nonumber\\
    &= n\, \mathrm{Tr}\!\left([F, q]\, q^{\,n-1}\right)
       = n\, \mathrm{Tr}(F q^n) - n\, \mathrm{Tr}(q F q^{\,n-1})
       = 0,
       \label{conservation:Qn}
\end{align}
where we have used the total symmetry of \(d^{abc\cdots}\) and the cyclicity of the trace.
Thus, the conservation of the classical Casimir  invariants follows automatically within the 
constrained Hamiltonian formalism. 
The remaining task is to construct a Hamiltonian that consistently preserves the phase-space 
constraints~(\ref{eq_shell}), (\ref{eq:PSp}), and~(\ref{eq_normal}) in 
the \((x^\mu, P^\mu, \omega^\mu, \pi^\mu)\) sector.

To derive covariant equations of motion within the Hamiltonian formalism, it is natural to begin with a Lorentz-scalar Hamiltonian. In Ref.~\cite{Heinz:1984my}, the Hamiltonian is expressed as
\begin{align}
H = \frac{1}{2m} \Big(P^2 - \frac{1}{2} g q^a F^a_{\mu\nu} S^{\mu\nu} - m^2 \Big),
\end{align}
which is obtained from the quadratic form of the QCD Dirac equation. Using this Hamiltonian along with the fundamental Poisson brackets of classical mechanics, as given in Eqs.~(\ref{eq_canon}), (\ref{eq:qP}), and (\ref{PBspin}), one can straightforwardly reproduce the covariant equations of motion derived in Ref.~\cite{Heinz:1984my}.
However, its consistency still relies on imposing conditions such as the spin condition $u_\mu S^{\mu \nu } = 0$ and the normalization condition $u^2 = 1$ with $P^\mu =mu^\mu $, together with an additional restriction $u^\mu\mathrm D_\mu (F_{\lambda\nu} S^{\lambda\nu}) = 0$ required to preserve the normalization condition. Thus, this formulation is not obtained within a closed Hamiltonian formalism. Therefore, in this study, we adopt the Hamiltonian formalism with constraints to derive the covariant equations of motion, providing a consistent and systematic framework.

Inspired by the Lorentz-scalar primary Hamiltonian for a U(1) gauge field in Ref.~\cite{Deriglazov_2014}, we construct the primary Hamiltonian for spin-$1/2$ colored particles (i.e., quarks) in a background non-Abelian Yang-Mills field as
\begin{align}
		H_P&=\dfrac{1}{2}g_1(P^2-2\mu gq^aF^{a}_{\mu\nu}\omega^\mu\pi^\nu-m^2)+\dfrac{1}{2}g_2(\pi^2-a_2)\nonumber
		\\&+\dfrac{1}{2}g_3(\omega^2-a_3)+g_5(P\pi)+\lambda_4(P\omega)+\lambda_6(\omega\pi)\nonumber
		\\&+\lambda_{g_i}\pi_{g_i}.
\end{align}
Here, $(g_i, \pi_{g_i})$ for $i=1,2,3,5$ are auxiliary canonical variable pairs associated with the constraints, satisfying the Poisson relations $\PB{g_i, \pi_{g_j}} = \delta_{ij}$, and $\lambda_{g_i}$ are the corresponding multipliers enforcing the primary constraints $\pi_{g_i}=0$. While $(g_i, \pi_{g_i})$ enlarge the phase space, the physical subspace is formed by $\zeta = (x^\mu, P^\mu, \omega^\mu, \pi^\nu, q^a)$, which contains all the physical degrees of freedom. The primary constraints are
\begin{equation}
	\label{eq:pc}
	\text{prcon}_i: \quad \pi_{g_i} = 0, \quad \omega \cdot \pi = 0, \quad P \cdot \omega = 0.
\end{equation}

\par The physical constraints from Sec.~\ref{se:cd} can be divided into two sets. The first set, denoted $T_i = 0$, reads
\begin{equation}
	\begin{split}
	T_1:-\dfrac{1}{2}(P^2-2\mu gq^aF^a_{\mu\nu}\omega^\mu\pi^\nu-m^2),\\
	T_2:-\dfrac{1}{2}(\pi^2-a_2),\quad T_3:-\dfrac{1}{2}(\omega^2-a_3),\\
	T_4:-P\omega,\quad T_5:-P\pi,\quad T_6:\omega\pi,
	\end{split}
\end{equation}
and the second set corresponds to the classical Casimirs, $Q_n = C$. Since Eq.~(\ref{conservation:Qn}) explicitly shows that the equations of motion automatically conserve the classical Casimirs, we can focus on applying the Dirac-Bergmann algorithm to the first set of constraints to derive consistent equations of motion. This demonstrates the overall consistency of the dynamics.

\par The constraints $T_4 = 0$ and $T_6 = 0$ arise from the primary constraints, whereas the other four constraints emerge at subsequent stages of the Dirac-Bergmann algorithm (see Appendix~\ref{ap:poisson}). After the full application of the algorithm, no new constraints appear in the physical phase space $\zeta$, but additional restrictions on the auxiliary variables $g_i$ and the Lagrange multipliers $\lambda_i$ are obtained. In particular, the auxiliary variables $g_1$ and $g_2$, together with their multipliers $\lambda_{g_1}$ and $\lambda_{g_2}$, remain undetermined. This corresponds to two first-class constraints: $\lambda_{g_1}$ and $g_1$ are associated with reparametrization invariance, while $\lambda_{g_2}$ and $g_2$ are associated with spin-plane invariance, as discussed in Ref.~\cite{Deriglazov_2014}.

\par Focusing on the physical subspace, we compute the Dirac brackets using only the constraints on $\zeta$, excluding the auxiliary variables. The matrix of Poisson brackets among these constraints is defined as
\begin{equation}
	\label{eq:cab}
	\mathcal{C}_{\alpha\beta} = \PB{T_\alpha, T_\beta}.
\end{equation}
The explicit form of $\mathcal{C}$ and its derivation are provided in Appendix~\ref{ap:poisson}. Its rank is $4$, indicating two first-class constraints and four second-class constraints. In constrained Hamiltonian formalism, each first-class constraint eliminates two degrees of freedom, and each second-class constraint eliminates one degree of freedom. Therefore, the total number of degrees of freedom associated with $(x^\mu, P^\mu, \omega^\mu, \pi^\nu)$ is
\begin{equation}
	2 \times 4 + 2 \times 4 - 2 \times 2 - 4 \times 1 = 6+2.
\end{equation}

\par The null vectors of the matrix $\mathcal{C}$ are found to be
\begin{subequations}
	\begin{gather}
		v_1=(\dfrac{m^2+(2\mu+1)c_3}{\mu b_2+(1-\mu)c_2},0,0,-\dfrac{\mu b_1+(1-\mu)c_1}{\mu b_2+(1-\mu) c_2},1,0),\\
		v_2=(0,\dfrac{a_3}{a_2},1,0,0,0),
	\end{gather}
\end{subequations}
where $b_1, b_2, c_1, c_2, c_3$ are shorthand notations given in Appendix \ref{ap:poisson}. Consequently, the first-class constraints are
\begin{subequations}
\label{first-class}
	\begin{gather}
	\chi_{1}=\dfrac{m^2+(2\mu+1)c_3}{\mu b_2+(1-\mu) c_2}T_1-\dfrac{\mu b_1+(1-\mu)c_1}{\mu b_2+(1-\mu)c_2}T_4+T_5,\\
	\chi_2=\dfrac{a_3}{a_2}T_2+T_3
	\end{gather}
\end{subequations}

For the second-class constraints, we take the following four:
\begin{equation}
\label{second-class}
	\begin{split}
	\xi_{i}&=\{T_3,T_4,T_5,T_6\}\\
	&=\{\dfrac{1}{2}(-\omega^2+a_3),-P\omega,-P\pi,\pi\omega\},
	\end{split}
\end{equation}
for which the matrix $\Delta_{\alpha\beta} = \PB{\xi_\alpha, \xi_\beta}$ is invertible. These second-class constraints are used to construct the Dirac bracket, with the resulting brackets on the basic phase-space variables summarized in Appendix \ref{ap:poisson}.

After setting all second-class constraints to zero in the extended Hamiltonian and redefining the Lagrange multipliers, the partially reduced Hamiltonian becomes
\begin{equation}
H_{PR} = \frac{1}{2}\lambda_1 \left(P^2 - 2\mu g q^a F^a_{\mu\nu}\omega^\mu\pi^\nu - m^2\right) + \frac{1}{2}\lambda_2 (\pi^2 - a_2).
\end{equation}
The equations of motion for the basic phase-space variables are then given by the Dirac brackets:
\begin{subequations}
	\label{eq:EOMb}
	\begin{gather}
	\dot{x}^\mu=\DB{x^\mu,H_{PR}},\quad \dot{P}^\mu=\DB{P^\mu,H_{PR}},\\
 \dot{\omega}^\mu=\DB{\omega^\mu,H_{PR}},\quad \dot{\pi}^\mu=\DB{\pi^\mu,H_{PR}}.
	\end{gather}
\end{subequations}

Explicitly, the equations of motion are
\begin{widetext}
\begin{subequations}
	\label{eq:EOM1}
	\begin{gather}
		\begin{split}
				\dot{x}^\mu&=\lambda_1P^\mu+\lambda_1\dfrac{1}{\Delta}g(\mu-1)q^aF^a_{\rho\sigma}P^\sigma(\omega^\rho\pi^\mu-\omega^\mu\pi^\rho)-\lambda_1\dfrac{1}{\Delta}\mu gg_{\alpha\beta}q^a(D^\alpha F^a_{\rho\sigma})\omega^\rho\pi^\sigma(\omega^\beta\pi^\mu-\pi^\beta\omega^\mu),
		\end{split}\\
		\begin{split}
			\dot{P}^\mu&=gq^ag^{\mu\rho}F^a_{\rho\sigma}\dot{x}^\sigma+\lambda_1\mu gq^a (D^\mu F^a_{\rho\sigma})\omega^\rho\pi^\sigma,
		\end{split}\\
		\begin{split}
			\dot{\omega}^\mu&=\lambda_2\pi^\mu-\dfrac{1}{\Delta}(\mu-1)\lambda_1gq^aF^a_{\rho\sigma}P^\rho\omega^\sigma P^\mu-\dfrac{1}{\Delta}\mu \lambda_1gg_{\alpha\beta}q^a(D^\alpha F^a_{\rho\sigma})P^\mu\pi^\sigma\omega^\rho\omega^\beta-\mu \lambda_1gq^aF^a_{\rho\sigma}g^{\mu\sigma}\omega^\rho,
		\end{split}\\
		\begin{split}
		\dot{\pi}^\mu&=-\dfrac{a_2\lambda_2\omega^\mu}{a_3}-\dfrac{1}{\Delta}(\mu-1)\lambda_1gq^aF^a_{\rho\sigma}P^\rho\pi^\sigma P^\mu-\dfrac{1}{\Delta}\mu \lambda_1gg_{\alpha\beta}q^a(D^\alpha F^a_{\rho\sigma})P^\mu\pi^\sigma\pi^\beta\omega^\rho+\mu \lambda_1 g q^aF^a_{\rho\sigma}g^{\mu\rho}\pi^\sigma,
	\end{split}\\
		\begin{split}
				\dot{q}^a&=-\lambda_1gf^{abc}q^cP^\mu A_\mu^b-\mu gf^{abc}q^c\lambda_1F^b_{\mu\nu}\omega^\mu\pi^\nu+\dfrac{1}{\Delta}(1-\mu)\lambda_1g^2q^cq^df^{abc}F^d_{\rho\sigma}A^{b}_{\mu}P^\rho(\pi^\sigma\omega^\mu-\omega^\sigma\pi^\mu)\\
			&-\dfrac{1}{\Delta}\mu \lambda_1g^2q^cq^df^{abc}(D_\nu F^d_{\rho\sigma})\omega^\rho\pi^\sigma A^{b}_{\mu}(\omega^\mu\pi^\nu-\omega^\nu\pi^\mu),
		\end{split}
	\end{gather}
\end{subequations}
\end{widetext}
where $\Delta = m^2 + (2\mu + 1) g q^a F^a_{\mu\nu} \omega^\mu \pi^\nu = m^2 + \frac{1}{4}(2\mu+1) g q^a F^a_{\mu\nu} S^{\mu\nu}$, and the covariant derivative is $q^a D_\mu F^{a}_{\rho\sigma} = q^a \partial_\mu F^{a}_{\rho\sigma} - g f^{abc} A^b_\mu q^c F^{a}_{\rho\sigma}$. The derivatives are taken with respect to the worldline parameter $\lambda$, i.e., $\dot{\zeta} = d\zeta/d\lambda$. It is noted that the above equations are well defined only in the regime $\Delta>0$. When the field becomes sufficiently strong that the spin-field interaction energy is comparable to the mass scale, the classical description is expected to break down.

It is apparent from Eq.~(\ref{eq:EOM1}) that all velocities are proportional to $\lambda_1$, reflecting the reparametrization invariance of the worldline. Another undetermined multiplier, $\lambda_2$, appears in $(\dot{\omega}, \dot{\pi})$. However, since it is associated only with reparametrization and spin-plane invariance, it does not affect the physical dynamics; specifically, the evolution of $S^{\mu\nu}$ is independent of $\lambda_2$.

Using Eq.~(\ref{eq:EOM1}) together with the definition (\ref{eq:St}), the equations of motion for the phase space variables can be written as
\begin{widetext}
\begin{subequations}
	\label{eq:EOM2}
	\begin{gather}
		\dot{x}^\mu=\lambda_1P^\mu+\lambda_1\dfrac{g}{2\Delta}(\mu-1)q^aF^a_{\rho\sigma}P^\sigma S^{\rho \mu}+\lambda_1\dfrac{g}{8\Delta}\mu q^a(D_\nu F^a_{\rho\sigma})S^{\rho\sigma}S^{\mu\nu},\\
		\dot{P}^\mu=gq^ag^{\mu\rho}F^a_{\rho\sigma}\dot{x}^\sigma+\dfrac{1}{4}\mu \lambda_1gq^a(D^\mu F^a_{\rho\sigma})S^{\rho\sigma},\\
		\dot{S}^{\mu\nu}=\dfrac{g}{\Delta}(\mu-1)\lambda_1q^aF^a_{\rho\sigma}P^\rho P^{[\mu}S^{\nu]\sigma}+\mu \lambda_1gq^aF^a_{\rho\sigma}g^{[\mu\rho}S^{\sigma \nu]} -\dfrac{g}{4\Delta}\lambda_1\mu q^a(D_\alpha F^a_{\rho\sigma})S^{\rho\sigma}P^{[\mu} S^{\alpha \nu]},\\
		\dot{q}^a=-gf^{abc}q^cA^b_{\mu}\dot{x}^\mu-\dfrac{1}{4}\mu \lambda_1gf^{abc}q^cF^b_{\mu\nu}S^{\mu\nu},
	\end{gather}
\end{subequations}
where \(A^{[\mu} B^{\nu]} \equiv A^\mu B^\nu - A^\nu B^\mu\) for arbitrary tensors \(A\) and \(B\); the antisymmetrization acts only on the indices \(\mu,\nu\), with all other indices left unchanged. 

These equations can be written more compactly using the matrix representation $q = q^a T^a$, $A_\mu = A_\mu^a T^a$, and $F_{\mu\nu} = F_{\mu\nu}^a T^a$:

\begin{subequations}
	\label{eq:EOMFM}
	\begin{gather}
		\dot{x}^\mu=\lambda_1P^\mu+\dfrac{g}{\Delta}\lambda_1S^{\mu\nu}\left\{\dfrac{\mu}{4}\tr{qD_\nu F_{\rho\sigma}}S^{\rho\sigma}-(\mu-1)\tr{qF_{\nu\rho}}P^\rho\right\},\\
		\dot{P}^\mu=2gg^{\mu\nu}\tr{qF_{\nu\rho}}\dot{x}^\rho+\dfrac{1}{2}\mu \lambda_1g\tr{qD^\mu F_{\rho\sigma}}S^{\rho\sigma},\\
		\dot{S}^{\mu\nu}=2\mu \lambda_1 g\tr{qF_{\rho\sigma}}g^{[\mu\rho}S^{\sigma \nu]}+\dfrac{2g}{\Delta}  \lambda_1P^{[\mu}S^{\nu]\alpha}\left\{\dfrac{\mu}{4}\tr{qD_\alpha F_{\rho\sigma}}S^{\rho\sigma}-(\mu-1)\tr{qF_{\alpha\rho}}P^\rho\right\},\\
			\dot{q}
			=ig[A_\mu,q]\dot{x}^\mu+i\dfrac{g}{4}\mu \lambda_1[F_{\mu\nu},q]S^{\mu\nu},
	\end{gather}
\end{subequations}
with the covariant derivative $D_\mu F_{\rho\sigma} = \partial_\mu F_{\rho\sigma} - i g [A_\mu, F_{\rho\sigma}]$. We note that, for the special choice $\mu=1$, $\lambda_1=1/m$ and $\dif x^\mu/\dif\tau=P^\mu/m$, and upon omitting the last two additional terms in the time derivative of the spin tensor in Eq.~(\ref{eq:EOMFM}), our equations of motion reduce to those of Ref.~\cite{Heinz:1984my}.

Equations (\ref{eq:EOM2}) can also be consistently transformed from the Frenkel-type spin tensor $S^{\mu\nu}$ to the BMT-type spin four-vector $S^\mu$ using Eq.~(\ref{eq:S}) \cite{Deriglazov_2014}. In terms of $S^\mu$, the equations of motion become
\begin{subequations}
	\begin{gather}
		\dot{x}^\mu=\lambda_1P^\mu+\lambda_1\dfrac{g}{2\Delta\sqrt{P^2}}(\mu-1)q^aF^a_{\rho\sigma}P^\sigma \epsilon^{\rho \mu\alpha\beta}P_\alpha S_\beta+\lambda_1\dfrac{g}{8\Delta P^2}\mu q^a(D_\nu F^a_{\rho\sigma})\epsilon^{\rho\sigma\alpha\beta}\epsilon^{\mu\nu\alpha'\beta'}P_\alpha S_\beta P_{\alpha'}S_{\beta'},\\
		\dot{P}^\mu=gq^ag^{\mu\nu}F^a_{\nu\rho}\dot{x}^\rho+\dfrac{g}{4\sqrt{P^2}}\mu \lambda_1q^a(D^\mu F^a_{\rho\sigma})\epsilon^{\rho\sigma\alpha\beta}P_\alpha S_\beta,\\
		\dot{S}^\mu=\mu \lambda_1gq^aF^{a,\mu\nu}S_{\nu}+\dfrac{1}{P^2}\mu \lambda_1gq^aF^{a,\rho\sigma}S_\rho P_\sigma P^\mu-\dfrac{1}{P^2}S^\nu\dot{P}_\nu P^\mu,\\
		\dot{q}^a=-gf^{abc}q^cA^b_\mu\dot{x}^\mu-\dfrac{g}{4\sqrt{P^2}}\mu \lambda_1 f^{abc}q^cF^b_{\mu\nu}\epsilon^{\mu\nu\rho\sigma}P_\rho S_\sigma.
	\end{gather}
\end{subequations}

Finally, these can also be expressed in a compact matrix form:
\begin{subequations}
	\label{eq:BMTs}
	\begin{gather}
		\dot{x}^\mu=\lambda_1P^\mu+\dfrac{g}{\Delta\sqrt{P^2}}\lambda_1\epsilon^{\mu\nu\alpha\beta}P_\alpha S_\beta\left\{\dfrac{\mu}{4}\tr{qD_\nu F_{\rho\sigma}}\epsilon^{\rho\sigma\alpha'\beta'}P_{\alpha'}S_{\beta'}-(\mu-1)\tr{qF_{\nu\rho}}P^\rho\right\},\\
		\dot{P}^\mu=2gg^{\mu\nu}\tr{qF_{\nu\rho}}\dot{x}^\rho+\dfrac{g}{2\sqrt{P^2}}\mu \lambda_1\epsilon^{\rho\sigma\alpha\beta}P_\alpha S_\beta\tr{qD^\mu F_{\rho\sigma}},\\
		\dot{S}^\mu=2\mu \lambda_1g\tr{qF^{\mu\nu}}S_\nu+\dfrac{1}{P^2}P^\mu \left\{2\mu \lambda_1g\tr{qF^{\rho\sigma}}S_\rho P_\sigma-S^\nu\dot{P}_\nu\right\},\\
		\dot{q}=ig[A_\mu,q]\dot{x}^\mu+i\dfrac{g}{4\sqrt{P^2}}\mu \lambda_1[F_{\mu\nu},q]\epsilon^{\mu\nu\alpha\beta}P_\alpha S_\beta.
	\end{gather}
\end{subequations}
\end{widetext}

\par Next, we explicitly verify gauge invariance of the equations of motion. Under a local non-Abelian gauge transformation, variables without color indices remain invariant, while the colored quantities transform as  
\begin{subequations}
\begin{align}
A'_\mu &= U A_\mu U^\dagger - \frac{i}{g} (\partial_\mu U) U^\dagger, \\
F'_{\mu\nu} &= U F_{\mu\nu} U^\dagger, \\
(D_\mu F_{\rho\sigma})' &= U D_\mu F_{\rho\sigma} U^\dagger, \\
q' &= U q U^\dagger,
\end{align}
\end{subequations}  
with \(U U^\dagger = I\). The first three equations of motion in Eqs.~(\ref{eq:EOMFM}) and (\ref{eq:BMTs}) are manifestly gauge invariant due to the trace operations.  

Without loss of generality, we focus on the last equation in Eq.~(\ref{eq:EOMFM}) for \(q\), note that  
\begin{equation}
\frac{d}{d\lambda}(q') = \frac{d}{d\lambda}(U q U^\dagger) = U \dot{q} U^\dagger + (\partial_\mu U \, q U^\dagger + U q \, \partial_\mu U^\dagger) \dot{x}^\mu,
\end{equation}  
and  
\begin{align}
([A_\mu, q])' &= U [A_\mu, q] U^\dagger - \frac{i}{g} [(\partial_\mu U) U^\dagger, U q U^\dagger] \nonumber \\
&= U [A_\mu, q] U^\dagger - \frac{i}{g} \left( (\partial_\mu U) q U^\dagger + U q \, \partial_\mu U^\dagger \right),
\end{align}  
where we used \((\partial_\mu U^\dagger) U + U^\dagger (\partial_\mu U) = 0\). Therefore, the full combination in the equation of motion  
\begin{equation}
\dot{q} - i g [A_\mu, q] \dot{x}^\mu - i \frac{g}{4} \mu \lambda_1 [F_{\mu\nu}, q] S^{\mu\nu}
\end{equation}  
transforms covariantly as  
\begin{equation}
(\cdot)' = U (\cdot) U^\dagger = 0,
\end{equation}  
showing gauge invariance.

\par Finally, we discuss the remaining undetermined Lagrange multiplier $\lambda_1$. In the constrained Hamiltonian formalism, such an undetermined multiplier is associated with a first-class constraint and reflects a gauge redundancy of the system. In the present case, this redundancy is the reparametrization invariance of the worldline, namely, the freedom to choose different parameters to describe the same physical trajectory. This redundancy can be removed by an appropriate gauge fixing, after which the explicit dependence of the equations of motion on $\lambda_1$ is eliminated. Common choices include proper time, light-cone time \(\lambda = x^+ = (x^0 + x^3)/\sqrt{2}\), or coordinate time \(\lambda = x^0\).  

For \(\lambda = x^0\), since \(dx^0/d\lambda = 1\), one can solve for \(\lambda_1\) as  
\begin{equation}
\label{eq:g1v}
\lambda_1 = \dfrac{1}{P^0 + \dfrac{g}{\Delta} S^{0i} \left\{ \dfrac{\mu}{4} \text{Tr}[q D_i (F S)] - (\mu-1) \text{Tr}[q F_{i\nu} P^\nu] \right\}},
\end{equation}  
with \(FS = F_{\rho\sigma} S^{\rho\sigma}\).  

This gauge fixing is equivalent to adding an additional constraint \(x^0 - \lambda = 0\) in the Hamiltonian formalism, effectively converting the first-class constraint \(\chi_1\) into a second-class constraint. Details are provided in Appendix~\ref{ap:td}.

\section{Discussion and summary}
\label{se:dis}

\par In our derived equations of motion for quarks, the quark trajectory is shown to be modified by both spin and the background Yang-Mills field. A similar phenomenon is known for electrons moving in an electromagnetic field, where such terms lead to a magnetic \emph{Zitterbewegung} of the trajectory \cite{Deriglazov_2014}: a relativistic classical particle with spin oscillates rapidly around the conventional classical trajectory. This effect also causes the velocity of the particle to be non-collinear with its kinetic momentum, a feature that has been observed in the Mathisson-Papapetrou-Dixon (MPD) equations for relativistic particles in curved spacetime \cite{Lukes_Gerakopoulos_2017}. 
\par Similarly, for a quark moving in a background Yang-Mills field, it experiences a force analogous to the Lorentz force combined with a spin-dependent Zeeman-like force. When we reformulate our equations in terms of the BMT-type spin vector (Eq.~(\ref{eq:BMTs})), the spin dynamics take a form analogous to the standard BMT theory \cite{Wen_2017}, confirming that the spin precession is influenced by both the non-Abelian Yang-Mills field and the quark's color charge. Moreover, the evolution of the classical color charge itself is shown to be modified by spin effects, consistent with earlier findings in \cite{Heinz:1984my}. This demonstrates that spin and color degrees of freedom are dynamically coupled in the presence of Yang-Mills fields.
\par In summary, by applying constrained Hamiltonian mechanics, we have derived a self-consistent set of equations of motion for a quark moving in a generic background Yang-Mills field. In this framework, the quark is modeled as a classical point like particle with position \(x^\mu\), kinetic momentum \(P^\mu\), spin tensor \(S^{\mu\nu}\), and classical color charge \(q\). The constraints inherent to the Hamiltonian formalism restrict the quark to move on the spin surface, even for \(g_S \neq 2\) and arbitrary background fields. We have explicitly verified that the equations of motion transform covariantly under non-Abelian gauge transformations and preserve the values of classical Casimir invariants, ensuring consistency with the underlying gauge symmetry.  
\par This framework provides a rigorous and self-consistent classical description of quark dynamics in a background Yang-Mills field, incorporating both spin and color degrees of freedom. It also offers a systematic approach to investigate spin-dependent phenomena in heavy-ion collisions, including the interplay between color fields and quark spin polarization, which is essential for understanding observed spin polarization, spin alignment, and the transport behavior of heavy quarks in the quark-gluon plasma.

\section*{ACKNOWLEDGEMENTS}
We thank Che Ming Ko for useful discussions and comments. This work was supported by the National Natural Science Foundation of China (NSFC) under Grant Nos. 12422508, 12375124, and the Science and Technology Commission of Shanghai Municipality under Grant No. 23JC1402700. Y.S. thanks the sponsorship from Yangyang Development Fund.

\appendix

\section{Dirac-Bergmann Algorithm and Dirac Brackets}
\label{ap:poisson}

Applying the Dirac-Bergmann algorithm, the self-consistency condition 
\begin{equation}
\dot{\text{prcon}}_i = \PB{\text{prcon}_i, H_P} \approx 0
\end{equation} 
generates the second-stage constraints:
\begin{widetext}
\begin{equation}
	\begin{split}
	\text{2ndcon}_i:\quad & -\frac{1}{2}\Big(P^2 - \frac{\mu g}{2} q^a F^{a,\mu\nu} S_{\mu\nu} - m^2\Big), \quad -\frac{1}{2}(\pi^2 - a_2), \quad -\frac{1}{2}(\omega^2 - a_3), \quad -P \cdot \pi,\\
	& g_2 \pi^2 - g_3 \omega^2, \quad 
	-g_5 \Big[m^2 + g (2 \mu +1) q^a F^a_{\mu\nu} \omega^\mu \pi^\nu\Big] + g_1 g (\mu-1) q^a F^a_{\mu\nu} P^\nu \omega^\mu - g_1 g \mu q^a (D^\mu F^a_{\rho\sigma}) g_{\mu\nu} \pi^\sigma \omega^\nu \omega^\rho.
	\end{split}
\end{equation}

In deriving these expressions, we used the following Poisson bracket relations:
\begin{subequations}
\begin{align}
\PB{P^\mu, P^\nu} &= g q^a F^{a,\mu\nu},\\
\PB{P^\mu, F^{a,\rho\sigma}} &= -\partial^\mu F^{a,\rho\sigma},\\
\PB{q^a, P^\mu} &= - g f^{abc} A^{b,\mu} q^c,
\end{align}
\end{subequations}
which follow from Eqs.~(\ref{eq:P}) and (\ref{eq:PBd}).

\par The second-stage constraint $\text{2ndcon}_5$ yields a relation between $g_2$ and $g_3$:
\begin{equation}
g_2 = \frac{a_3}{a_2} g_3,
\end{equation}
while $\text{2ndcon}_6$ provides a relation between $g_5$ and $g_1$:
\begin{equation}
\label{eq:g6}
g_5 = \frac{g_1 g (\mu-1) q^a F^a_{\mu\nu} P^\nu \omega^\mu - g_1 g \mu q^a (D^\mu F^a_{\rho\sigma}) g_{\mu\nu} \pi^\sigma \omega^\nu \omega^\rho}{m^2 + (2 \mu + 1) gq^a F^a_{\mu\nu} \omega^\mu \pi^\nu}.
\end{equation}

\par Proceeding to the third stage of the Dirac-Bergmann algorithm, we apply 
\begin{equation}
\dot{\text{2ndcon}}_i = \PB{\text{2ndcon}_i, H_P} \approx 0,
\end{equation} 
which gives
\begin{equation}
	\begin{split}
	\dot{\text{2ndcon}}_i:\quad & g_5 (\mu-1) g q^a F^a_{\mu\nu} P^\mu \pi^\nu + g_5 g \mu q^a (D^\rho F^a_{\mu\nu}) g_{\rho\sigma} \pi^\nu \pi^\sigma \omega^\mu - \lambda_4 (\mu-1) g q^a F^a_{\mu\nu} P^\nu \omega^\mu + \lambda_4 g \mu q^a (D^\rho F^a_{\mu\nu}) g_{\rho\sigma} \pi^\nu \omega^\mu \omega^\sigma,\\
	& \lambda_6 \pi^2, \quad -\lambda_6 \omega^2, \quad \lambda_4 \big[m^2 + g (2 \mu +1) q^a F^a_{\mu\nu} \omega^\mu \pi^\nu\big] + g_1 g (\mu-1) q^a F^a_{\mu\nu} P^\nu \pi^\mu - g_1 g \mu q^a (D^\mu F^a_{\rho\sigma}) g_{\mu\nu} \pi^\nu \pi^\sigma \omega^\rho,\\
	& -2 g_2 \lambda_6 \pi^2 - 2 g_3 \lambda_6 \omega^2 + \lambda_{g_2} \pi^2 - \lambda_{g_3} \omega^2, \quad f_1(\lambda_{g_i}, \lambda_4, \lambda_6).
	\end{split}
\end{equation}

From $\dot{\text{2ndcon}}_4 \approx 0$, we immediately obtain $\lambda_6 = 0$, and the relation between $\lambda_4$ and $g_1$:
\begin{equation}
\label{eq:g5}
\lambda_4 = - \frac{g_1 g (\mu-1) q^a F^a_{\mu\nu} P^\nu \pi^\mu - g_1 g \mu q^a (D^\mu F^a_{\rho\sigma}) g_{\mu\nu} \pi^\nu \pi^\sigma \omega^\rho}{m^2 + (2 \mu +1) gq^a F^a_{\mu\nu} \omega^\mu \pi^\nu}.
\end{equation}

\par Inserting Eqs.~(\ref{eq:g6}) and (\ref{eq:g5}) into $\dot{\text{2ndcon}}_1$, one finds that $\dot{\text{2ndcon}}_1$ is not independent of $\dot{\text{2ndcon}}_4$ and $\text{2ndcon}_6$. Moreover, all $\dot{\text{2ndcon}}_i$ for $i=1,\dots,6$ only impose restrictions on Lagrange multipliers and do not generate new constraints. Therefore, the Dirac-Bergmann algorithm terminates at the third stage.

\par Collecting all phase-space constraints as $\text{con}_i$, we have
\begin{equation}
	\begin{split}
	\text{con}_i = \big\{ &\pi_{g_1}, \pi_{g_2}, \pi_{g_3}, \pi_{g_5}, -\frac{1}{2}(P^2 - 2 g \mu q^a F^a_{\mu\nu} \omega^\mu \pi^\nu - m^2), \frac{1}{2}(-\pi^2 + a_2), \frac{1}{2}(-\omega^2 + a_3), -P\cdot\omega, -P\cdot\pi, \pi\cdot\omega, a_2 g_2 - a_3 g_3, \\
	& - g_5 \big[m^2 + g (2\mu+1) q^a F^a_{\mu\nu} \omega^\mu \pi^\nu\big] + g_1 g (\mu-1) q^a F^a_{\mu\nu} P^\nu \omega^\mu - g_1 g \mu q^a (D^\mu F^a_{\rho\sigma}) g_{\mu\nu} \pi^\sigma \omega^\nu \omega^\rho \big\}.
	\end{split}
\end{equation}

\par As shown in Ref.~\cite{Deriglazov_2014}, the dynamics can be projected onto the physical sub-phase space $\zeta = (x^\mu, P^\mu, \omega^\mu, \pi^\mu, q^a)$, leading to the physical constraints:
\begin{equation}
	T_1 = -\frac{1}{2}(P^2 - 2 \mu g q^a F^a_{\mu\nu} \omega^\mu \pi^\nu - m^2), \quad
	T_2 = -\frac{1}{2}(\pi^2 - a_2), \quad
	T_3 = -\frac{1}{2}(\omega^2 - a_3), \quad
	T_4 = -P\cdot\omega, \quad
	T_5 = -P\cdot\pi, \quad
	T_6 = \pi \cdot \omega.
\end{equation}

\par Denoting the Poisson brackets among these constraints as
\begin{equation}
\mathcal{C}_{\alpha\beta} = \PB{T_\alpha, T_\beta},
\end{equation}
the explicit form of $\mathcal{C}_{\alpha\beta}$ in terms of phase-space variables is
\begin{equation}
\mathcal{C}_{\alpha\beta} =
\begin{pmatrix}
0 & 0 & 0 & (1-\mu)c_2 + \mu b_2 & (1-\mu)c_1 + \mu b_1 & 0 \\
0 & 0 & 0 & 0 & 0 & a_2 \\
0 & 0 & 0 & 0 & 0 & -a_3 \\
(\mu-1)c_2 - \mu b_2 & 0 & 0 & 0 & (2\mu+1)c_3 + m^2 & 0 \\
(\mu-1)c_1 - \mu b_1 & 0 & 0 & -(2\mu+1)c_3 - m^2 & 0 & 0 \\
0 & -a_2 & a_3 & 0 & 0 & 0
\end{pmatrix},
\end{equation}
with
\begin{subequations}
\begin{align}
c_1 &= g q^a F^a_{\mu\nu} P^\mu \pi^\nu, \quad c_2 = g q^a F^a_{\mu\nu} P^\mu \omega^\nu, \quad c_3 = g q^a F^a_{\mu\nu} \omega^\mu \pi^\nu,\\
b_1 &= - g_{\rho\sigma} g q^a (\partial^\rho F^a_{\mu\nu}) \pi^\nu \omega^\mu \pi^\sigma + g^2 f^{abc} q^c F^a_{\mu\nu} \omega^\mu \pi^\nu A^{b,\rho} \pi_\rho,\\
b_2 &= - g_{\rho\sigma} g q^a (\partial^\rho F^a_{\mu\nu}) \pi^\nu \omega^\mu \omega^\sigma + g^2 f^{abc} q^c F^a_{\mu\nu} \omega^\mu \pi^\nu A^{b,\rho} \omega_\rho.
\end{align}
\end{subequations}

\par With the first-class and second-class constraints identified in Eqs.~(\ref{first-class}) and (\ref{second-class}), respectively, we can construct the Dirac bracket among the fundamental variables. In the position-momentum sector, the Dirac brackets read
\begin{subequations}
\begin{align}
\DB{x^\mu, x^\nu} &= \frac{\pi^\mu \omega^\nu - \omega^\mu \pi^\nu}{\Delta},\\
\DB{P^\mu, P^\nu} &= g q^a g^{\mu\rho} g^{\nu\sigma} F^a_{\rho\sigma} + \frac{1}{\Delta} g^2 q^a q^b F^a_{\rho\sigma} F^b_{\alpha\beta} (g^{\mu\rho} g^{\nu\beta} \pi^\alpha \omega^\sigma - g^{\mu\alpha} g^{\nu\sigma} \pi^\beta \omega^\rho),\\
\DB{x^\mu, P^\nu} &= g^{\mu\nu} + \frac{1}{\Delta} g q^a F^a_{\rho\sigma} (\pi^\rho g^{\nu\sigma} \omega^\mu - \pi^\mu g^{\nu\sigma} \omega^\rho),\\
\DB{x^\mu, F^a_{\rho\sigma}} &= \frac{1}{\Delta} (\partial_\nu F^a_{\rho\sigma}) (\pi^\mu \omega^\nu - \pi^\nu \omega^\mu),\\
\DB{P^\mu, F^a_{\rho\sigma}} &= -\partial^\mu F^a_{\rho\sigma} + \frac{g}{\Delta} (\partial_\nu F^a_{\rho\sigma}) F^b_{\alpha\beta} g^{\mu\alpha} (\pi^\beta \omega^\nu - \pi^\nu \omega^\beta),
\end{align}
\end{subequations}
where $\Delta = m^2 + gq^a (2\mu+1) F^a_{\mu\nu} \omega^\mu \pi^\nu = P^2 + \frac{1}{4} gq^a F^a_{\mu\nu} S^{\mu\nu}$.

The Dirac brackets in the spin and color sectors are provided in full in Eqs.~(\ref{spinDB}) and (\ref{colorDB}), completing the algebraic structure of the constrained phase space. In the spin sector, they read
\begin{subequations}
\label{spinDB}
	\begin{gather}
		\DB{\omega^\mu,\omega^\nu}=0,\quad 
		\DB{\omega^\mu,\pi^\nu}=g^{\mu\nu}-\dfrac{P^\mu P^\nu}{\Delta}-\dfrac{\omega^\mu\omega^\nu}{a_3},\quad 
		\DB{\pi^\mu,\pi^\nu}=\dfrac{\pi^\mu\omega^\nu-\omega^\mu\pi^\nu}{a_3},\\
		\DB{x^\mu,\pi^\nu}=-\dfrac{1}{\Delta}P^\nu \pi^\mu,\quad 
		\DB{P^\mu,\pi^\nu}=-\dfrac{g}{\Delta}q^aF^a_{\rho\sigma}P^\nu g^{\mu\rho}\pi^\sigma,\\
		\DB{x^\mu,\omega^\nu}=-\dfrac{1}{\Delta}P^\nu\omega^\mu,\quad 
		\DB{P^\mu,\omega^\nu}=-\dfrac{g}{\Delta}q^aF^a_{\rho\sigma}P^\nu g^{\mu\rho}\omega^\nu,\\
		\DB{\omega^\mu,F^a_{\rho\sigma}}=\dfrac{1}{\Delta}(\partial_\nu F^a_{\rho\sigma})P^\mu\omega^\nu,\quad 
		\DB{\pi^\mu,F^a_{\rho\sigma}}=\dfrac{1}{\Delta}(\partial_\nu F^a_{\rho\sigma})P^\mu\pi^\nu.
	\end{gather}
\end{subequations}

In the color sector, the Dirac brackets take the form
\begin{subequations}
\label{colorDB}
	\begin{gather}
		\DB{q^a,q^b}=f^{abc}q^c+\dfrac{1}{\Delta}g^2 f^{acd}f^{bef}q^dq^fA^c_\mu A^e_\nu(\pi^\mu\omega^\nu-\pi^\nu\omega^\mu),\\
		\DB{q^a,x^\mu}=\dfrac{g}{\Delta}A^b_\nu f^{abc}q^c(\pi^\mu\omega^\nu-\omega^\mu\pi^\nu),\\
		\DB{q^a,P^\mu}=-gf^{abc}q^cA^{b,\mu}+\dfrac{g^2}{\Delta}f^{abc}A^{b}_\nu F^d_{\alpha\beta}q^c q^dg^{\mu\beta}(\pi^\nu\omega^\alpha-\omega^\nu\pi^\alpha),\\
		\DB{q^a,\omega^\mu}=\dfrac{1}{\Delta}gf^{abc}q^cA^{b}_\nu P^\mu\omega^\nu,\\
		\DB{q^a,\pi^\mu}=\dfrac{1}{\Delta}gf^{abc}q^cA^{b}_\nu P^\mu\pi^\nu,\\
		\DB{q^a,F^b_{\mu\nu}}=\dfrac{1}{\Delta}gf^{adc}q^cA^{d}_\sigma(\partial_\rho F^b_{\mu\nu})(\pi^\rho\omega^\sigma-\omega^\rho\pi^\sigma).
	\end{gather}
\end{subequations}
\end{widetext}

\section{Hamiltonian formalism with time-dependent constraints}
\label{ap:td}

In this Appendix, we briefly summarize the Hamiltonian mechanics for systems with time-dependent constraints. If a function $F$ explicitly depends on the worldline parameter $\lambda$, its evolution is given by
\begin{equation}
	\dot{F} = \PB{F,H} + \frac{\partial F}{\partial \lambda}.
\end{equation}

When a constraint explicitly depends on $\lambda$, the Dirac-Bergmann self-consistency condition generalizes to
\begin{equation}
	\PB{\text{con}_i,H_P} + \frac{\partial \text{con}_i}{\partial \lambda} \approx 0.
\end{equation}

In our case, imposing a coordinate gauge constraint
\begin{equation}
	T_g: x^0 - \lambda = 0
\end{equation}
requires
\begin{equation}
	\PB{T_g,H_P} + \frac{\partial T_g}{\partial \lambda} = 0,
\end{equation}
which leads to a relation among $g_1$, $\lambda_4$, and $g_5$:
\begin{equation}
	\label{eq:g156}
	g_1 P^0 + \lambda_4 \omega^0 + g_5 \pi^0 - 1 = 0.
\end{equation}

Inserting Eqs.~(\ref{eq:g5}), (\ref{eq:g6}), and (\ref{eq:St}) into Eq.~(\ref{eq:g156}) determines the value of $g_1$ as
\begin{equation}
	\label{eq:g1}
	g_1 = \frac{1}{P^0 + \dfrac{g}{\Delta} S^{0i} \left( \dfrac{\mu}{4} \tr{q D_i F_{\rho\sigma} S^{\rho\sigma}} - (\mu-1) \tr{q F_{i\nu} P^\nu} \right)}.
\end{equation}

This expression is consistent with the result obtained via the gauge-fixing procedure in Eq.~(\ref{eq:g1v}). 

Noting that
\begin{equation}
	\PB{T_g, \chi_1} \neq 0,
\end{equation}
we see that the previously first-class constraint $\chi_1$ is converted into a second-class constraint upon imposing the gauge condition $T_g = 0$.

\nocite{}
\bibliography{references}

\end{document}